\documentclass[prb,twocolumn,showpacs,showkeys,amssymb]{revtex4}

\usepackage{graphicx}
\usepackage{dcolumn}
\usepackage{bm}

\begin{document}

\title{Effects of Coulomb interactions on the splitting of
 luminescence lines}
\author{Boris A. Rodriguez$^a$ and Augusto Gonzalez$^{a,b}$}
\affiliation{$^a$Instituto de Fisica, Universidad de Antioquia,
 AA 1226, Medellin, Colombia\\
 $^b$Instituto de Cibernetica, Matematica y Fisica, Calle E 309, Vedado,
 Ciudad Habana, Cuba.}

\begin{abstract}
We study the splitting between the right-hand and left-hand circularly polarized
luminescence lines in a quantum dot under relatively weak confinement regime and resonant high-power excitation. When the dot is populated with an even number of electron-hole pairs (biexciton and higher excitations), the splitting measures basically the Zeeman energy. However, in the odd number of pairs case, we have, in addition to the Zeeman and Overhauser shifts, a contribution to the splitting coming from Coulomb interactions. This contribution
is of the order of a few meV, and shows distinct signatures of shell-filling in the
quantum dot.
\end{abstract}


\maketitle

\section{Introduction}

This paper is aimed at a theoretical description of the splitting of
luminescence lines coming from the deexcitation of a quantum dot in the
presence of a magnetic field. There are a few factors contributing to
this splitting.

First, there is a contribution coming from the Zeeman energy, which is
proportional to the magnetic field, $B$. As luminescence lines arise
from exciton recombination, both the electron and hole Lande factors
enter the expression for the Zeeman shift. In a nanostructure, Lande
factors exhibit a dependence not only on the semiconductor band structure,
but also on the geometry. In the recent past, for example, extensive studies
on the dependence on well width \cite{Lande} have been conduced. For
magnetic fields between 1 and 3 Teslas, the typical values we will
encounter for the Zeeman shifts are 0.05 - 0.2 meV.

A second contribution to the splitting is the so called Overhauser shift
\cite{Overhauser}. It is originated in the hyperfine interaction between the
electron and nuclear spins. The electron motion around the nanostructure
induces a nuclear spin polarization. Once the nuclear polarization is set
up, the interaction of the electron with the thousands of nuclei conforming
the dot leads to a measurable shift in the luminescence lines. In a first
approximation, the Overhauser shift does not depend on the magnetic field
nor on the quantum dot size, but increases up to a saturation value with the
excitation power. The latter fact can be understood in terms of the increase
of the nuclear polarization as the rate of electron-hole pair creation is
raised. Typical values of the Overhauser shift are around 0.1 meV. Recent
measurements of this contribution were done on quantum dots formed from
width fluctuations in a well\cite{main2}, and on self-assembled quantum
dots\cite{self}.

There is a third contribution to the splitting, which shall appear when
the excitation power is high enough to create multiexcitonic complexes.
Recombination of one electron-hole pair from a complex involves Coulomb
interactions in a nontrivial way. The emission of a right-hand circularly
polarized photon proves to be not equivalent to the emission of a
left-hand circularly polarized one. Typical values for this shift are
around 3 meV, i.e., the values of energy separations between electronic
states in the complex. These Coulomb interaction effects have not been
studied experimentally nor theoretically so far. They constitute the
focus of our attention.

The plan of the paper is as follows. In Sec. \ref{sec2}, we present a
brief resume of known ideas explaining the experimental results of
Refs. [\onlinecite{main2},\onlinecite{self}], which correspond to low
excitation power and independent exciton recombination. This section
is intended as an introduction to Sec. \ref{sec3}, where higher
excitations are considered. Multiexcitonic complexes are treated in
the framework of a corrected BCS scheme, which takes a proper account
of mean field Coulomb interactions and Fermi statistics. We compute
the position and intensities of luminescence lines coming from the
decay of systems with up to 21 pairs. The main results are the absence
of Overhauser or Coulomb interaction effects in complexes with an even
number of pairs, and the shell filling effects in the shifts due to
Coulomb interactions for systems with an odd number of pairs. Concluding
remarks are given in the last section.

\section{The independent-exciton approximation to the $\sigma_+$ -
$\sigma_-$ splitting}
\label{sec2}

\begin{figure}[t]
\begin{center}
\includegraphics[width=.9\linewidth,angle=0]{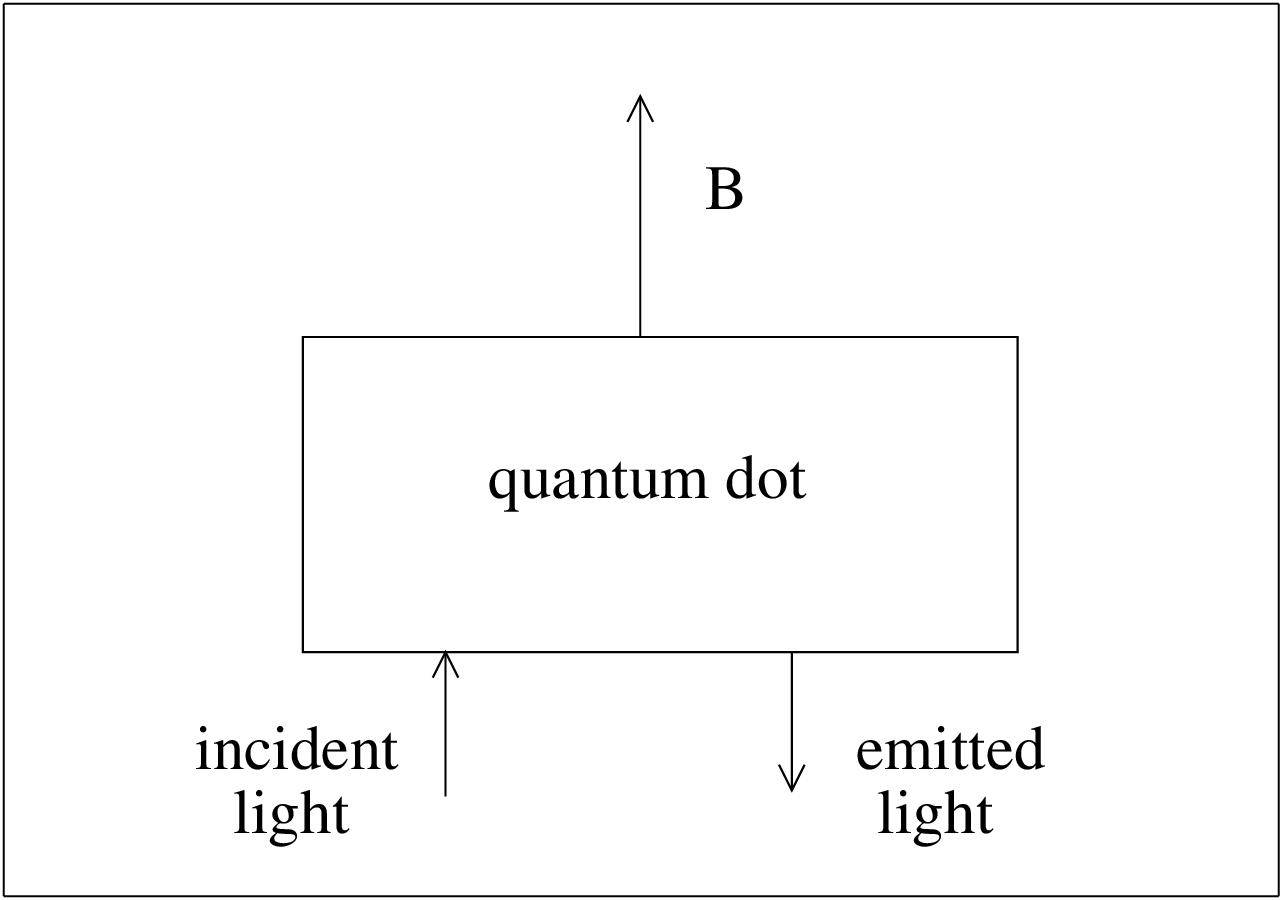}
\caption{\label{fig1} Geometry of the experiment.}
\end{center}
\end{figure}

The experiments \cite{main,main2,self} are usually performed in the
so called Faraday configuration, where the excitation laser beam
is parallel to the applied magnetic field, and backscattering geometry.
A schematic view is presented in Fig. \ref{fig1}.

The model parameters are choosen to fit (at least qualitatively) the
experimental setup described in paper [\onlinecite{main}]. A quasi
twodimensional quantum dot is formed from width fluctuations in a 4.2 nm-wide GaAs-AlGaAs
quantum well. The dot area is around $10^3$ nm$^2$. Taking into account
that the GaAs lattice constant is $\sim$ .3 nm, the number of nuclei
in the dot is $\sim 10^5$.

\begin{figure}[t]
\begin{center}
\includegraphics[width=.8\linewidth,angle=0]{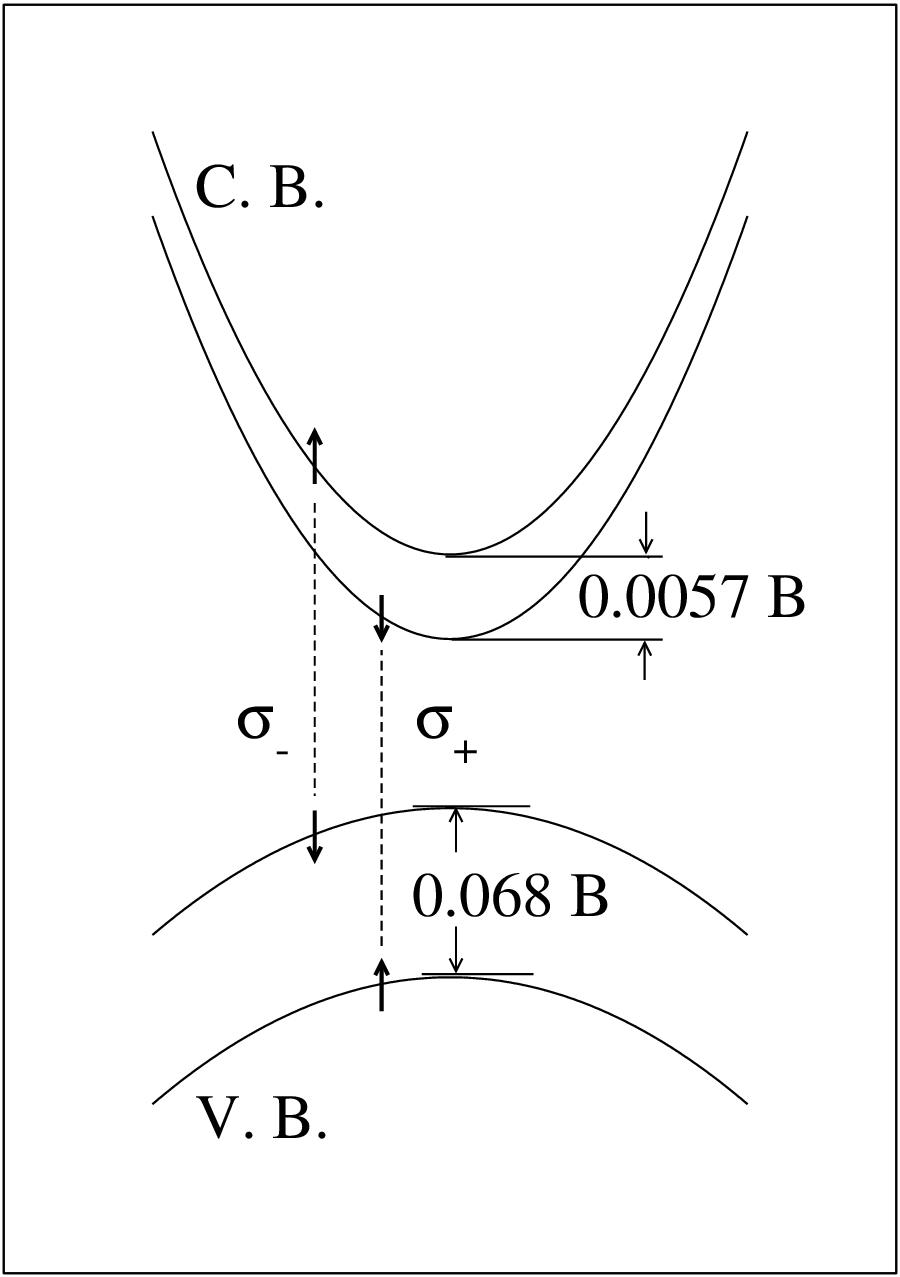}
\caption{\label{fig2} Schematics of the assumed band sructure
 in the dot. The electron-hole pairs created in the absorption
 of $\sigma_+$ and $\sigma_-$ photons are drawn.}
\end{center}
\end{figure}

A simplified two-band structure for GaAs is assumed. A schematics of
the band structure is drawn in Fig. \ref{fig2}, where electron -
hole coupling for each light polarization, and Zeeman splitting in
each sub-band are indicated. The effective electron and heavy-hole
Lande factors for the 4.2 nm-wide quantum well are expected to be
$g_e\approx 0.1$, $g_h\approx -1.2$. \cite{Lande} The Zeeman energy
of an electron-hole pair is computed from:

\begin{equation}
\varepsilon_{Zeeman}=\mu_B B (g_e S_e - g_h S_h)=\mu_B B (g_e + g_h) S_e,
\label{eq1}
\end{equation}

\noindent
where $\mu_B=0.057$ meV/Teslas is the atomic Bohr magneton, and $S_e,
S_h=\pm 1/2$ are the electron and hole spin projections over $B$,
which acts along the dot normal. $g_e+g_h\approx -1.1$ is the exciton
Lande factor.

The quantum dot is resonantly excited, i.e., the laser excitation
energy is only a few meV above the luminescence lines. Consequently,
in luminescence there is memory of how the dot is pumped. Notice that
the sense of rotation of the electric field is inverted for the
emitted light. This means that by pumping with $\sigma_+$ light,
for example, we are in fact populating the $\sigma_-$ mode for
backward emission.

As mentioned before, there are around $10^5$ nuclei in our GaAs dot.
Half of them are $^{69}$Ga and $^{71}$Ga nuclei with relative
abundances 60\% and 40\% respectively. The second half corresponds
to $^{75}$As nuclei. For all these nuclei, the total spin is
$I=3/2$. Their magnetic moments (in units of the nuclear magneton,
$\mu_n$) are, respectively, 2.017, 2.562 and 1.439. \cite{radzig}

The hyperfine interaction between one conduction band electron and nuclei
in the quantum dot is described by the Hamiltonian:

\begin{equation}
{\cal H}_{en}=\alpha\sum_{n}\vec S_e\cdot\vec I_n|\psi_e(\vec r_n)|^2,
\label{eq2}
\end{equation}

\noindent
where $\alpha=(2/3)\mu_0 g_e g_n \mu_B \mu_n$, and $\mu_0$ is the
magnetic permitivity of vacuum. $\psi_e$ is the electron wave function,
which contains both an envelope (orbital) part and the Bloch functions,
which are $s$-functions. $\psi_e$ is evaluated at the positions of the
nuclei, i.e. the origin of each cell. For Bloch $s$-functions, $\phi$,
we have $|\phi(0)|^2\approx Z^3/(a_B n)^3$, where $Z$ is the atomic number
(31 for Ga and 33 for As), $a_B$ is the Bohr radius, and $n=4$ corresponds
to the 4s electrons conforming the conduction band. The fact that
the orbital wave function is extended all over the dot (meaning that
the square of the orbital wave function is proportional to $1/N_n$, where
$N_n$ is the number of nuclei in the dot) make the hyperfine contribution
to the energy, Eq. (\ref{eq2}), roughly independent of the
dot size. In a mean field approximation, $\langle \vec I\rangle$ is
oriented along $B$, and we get for the interaction energy of one electron
with the $10^5$ nuclei in the dot:

\begin{equation}
\varepsilon_{hyperfine}= g_{en}\langle I\rangle S_e.
\label{eq3}
\end{equation}

\noindent
The $g_{en}$ constant for bulk GaAs was found to be around -0.090 meV.
\cite{gen} A similar value was observed in the quantum dots studied
in Refs. \onlinecite{main} and \onlinecite{main2}.

Finally, there is a contribution to the exciton energy coming from
the electron-hole exchange interaction in a deformed quantum dot. For
the quantum dot under study \cite{exch}, we have

\begin{equation}
\varepsilon_{\pm}^{(exch)}\approx \mp ~ 0.012 meV.
\label{eq4}
\end{equation}

With the help of Eqs. (\ref{eq1},\ref{eq3},\ref{eq4}), and the
independent exciton approximation, we write the single pair energy
in the form:

\begin{equation}
\varepsilon_{\pm}=\varepsilon_X + \varepsilon_{\pm}^{(exch)}+
 \mu_B B (g_e + g_h) S_e + g_{en}\langle I\rangle S_e,
\label{eq5}
\end{equation}

\noindent
where the + index corresponds to $S_e=1/2$ ($\sigma_+$ polarization of
the emitted light), and the - index to $S_e=-1/2$ ($\sigma_-$ polarization
of the emitted light). $\varepsilon_X$ is the contribution of the
electron-hole Coulomb attraction.
$\varepsilon_{\pm}$ gives, respectively, the position of the $\sigma_+$
and $\sigma_-$ luminescence lines. The shift between them is

\begin{equation}
\Delta\varepsilon=\varepsilon_--\varepsilon_+=\varepsilon^{(exch)}
- \mu_B B (g_e + g_h)-g_{en}\langle I\rangle.
\end{equation}

\noindent
It depends on the excitonic populations, $N_+$, $N_-$, through
$\langle I\rangle$. The Overhauser shift is, by definition,

\begin{equation}
\Delta\varepsilon_{Overhauser}=-g_{en}\langle I\rangle.
\end{equation}

\noindent
It can be measured by turning off the contribution of each nuclear
species to the mean nuclear spin with the help of microwave
radiation sintonized to the frequency of the corresponding nuclear
magnetic resonance \cite{nmr}. Under $\sigma_+$ pumping, the electron
spin polarization is $\langle S_e\rangle < 0$, and the mean nuclear spin is
$\langle I\rangle > 0$ in order to minimize the hyperfine interaction
energy. This means that the Overhauser shift is added to the Zeeman
splitting. Under $\sigma_-$ pumping, on the contrary, the Overhauser
shift is substracted from the Zeeman splitting.

\section{Account of many-particle Coulomb interactions}
\label{sec3}

In the independent-exciton approximation, Eq. (\ref{eq5}), the
dependence on pumping comes only from $\langle I\rangle$. In the experiments
reported in Refs. \onlinecite{main2, self}, the mean exciton number in the dot is less
than one. In paper \onlinecite{main}, there is a study of the
dependence of $\Delta\varepsilon$ on the excitation power, but what is indeed
observed in the single-exciton line is the saturation of the nuclear polarization, $\langle I\rangle$, as the occupation of the exciton state increases.

In the present section, we shall discuss the more general situation,
where the excitation power is enough to create a few pairs in the dot.
Many-body Coulomb interactions may introduce an additional dependence of
$\Delta\varepsilon$ on $N_+$ and $N_-$. The energy of a system
with $N_+, N_-$ pairs is written as:

\begin{eqnarray}
E(N_+,N_-)&=&E_{Coul}(N_+,N_-)\nonumber\\
&+& \mu_B B (g_e + g_h) (N_+-N_-)/2\nonumber\\
&+& g_{en}\langle I\rangle (N_+-N_-)/2,
\end{eqnarray}

\noindent
where $E_{Coul}$ accounts for Coulomb interactions among particles.
The positions of the $\sigma_+$ and $\sigma_-$ lines are obtained
from:

\begin{eqnarray}
\varepsilon_+&=&E_{Coul}(N_+,N_-)-E_{Coul}(N_+-1,N_-)\nonumber\\
 &+& \mu_B B (g_e + g_h)/2 + g_{en}\langle I\rangle/2,
\end{eqnarray}

\begin{eqnarray}
\varepsilon_-&=&E_{Coul}(N_+,N_-)-E_{Coul}(N_+,N_--1)\nonumber\\
 &-& \mu_B B (g_e + g_h)/2 - g_{en}\langle I\rangle/2.
\end{eqnarray}

\noindent
Coulomb interactions introduce a contribution to the shift:

\begin{equation}
\Delta\varepsilon_{Coul}=E_{Coul}(N_+-1,N_-)-E_{Coul}(N_+,N_--1).
\label{eq11}
\end{equation}

\noindent
It is reasonable to accept that $E_{Coul}$ does not vary when $N_+$ and $N_-$ are
interchanged: $E_{Coul}(N_+,N_-)=E_{Coul}(N_-,N_+)$.

In order to compute $E_{Coul}$, we use a model of parabolic dot with
harmonic confinement, $\hbar\omega\approx 10$ meV. This value of
$\omega$ is chosen to reproduce the exciton diamagnetic shift of
0.025 meV/Teslas$^2$ reported in Refs. [\onlinecite{main,main2}].
The corresponding oscillator
lenght is around 12 nm, a value similar to the dot characteristic
dimensions. Effective masses $m_e=0.067 ~ m_0$,
$m_h=0.15 ~ m_0$, and relative dielectric constant $\kappa=12.5$ are
used. Electron-hole pairing is accounted for by means of a BCS wave
function. Notice that the Bohr radius for GaAs is aroud 7 nm. This
means that the area occupied by an exciton is around 150 nm$^2$.
In our $10^3$ nm$^2$ dot, 7 excitons are already closed packed,
and the effects of Fermi statistics should be important. These
effects and the mean field Coulomb interactions are correctly
described by the BCS function.

Fluctuations in the particle number, not conserved by the BCS
function, are important for a system with around 10 pairs. Thus,
the BCS function should be projected onto a subspace with fixed
particle number. The projection scheme we use is the so called
Lipkin-Nogami scheme \cite{LN,PRB}. In the present situation, we
shall use a Lipkin-Nogami scheme with two conserved charges,
$N_+$ and $N_-$. As this situation is not common in the literature,
we give in the Appendix a brief description of the method and
the explicit equations to be used.

Let us first discuss the simplest case beyond the independent exciton
approximation: the $\sigma_+ - \sigma_-$ splitting in the biexciton
line.

The lowest biexciton state has $(N_+,N_-)=(1,1)$. The first state with
$N_+$ or $N_-$ equal to 2 has excitation energy higher than 8 meV (with
our model parameters). Under quasiresonant excitation and temperatures
below 4 K, only the ground state will be populated with high probability.

The Coulomb contribution to the splitting, Eq. (\ref{eq11}), is thus zero
for the biexciton lines. It is not difficult to see that the Overhauser shift
is also zero or, at least, much smaller than the shift for the exciton.
Indeed, an intuitive argument suggests that, as the spin polarization is zero
in the ground state, the mean nuclear spin will be zero too. In fact, the
situation is a bit more complicated. One can guess that, as each excitation has
a certain weight in the density matrix, exciton states, including the long-lived
dark states, play an important role in determining the mean nuclear spin \cite{main2}.
Nevertheless, a lower value for $\langle I\rangle$ is expected and, consequently,
a lower shift due to the Overhauser effect.

There is, nevertheless, a very important effect of Coulomb interactions consisting in
the concentration of the luminescence in a single, coherent, line, as will become clear below.

The conclusion is that the biexciton lines exhibit a weaken Overhauser effect,
and no shift coming from Coulomb interactions among particles. This conclusion
can be extended to any states with $N_+=N_-$, i.e., when there is an even number
of pairs in the dot.

\begin{figure}[t]
\begin{center}
\includegraphics[width=.99\linewidth,angle=0]{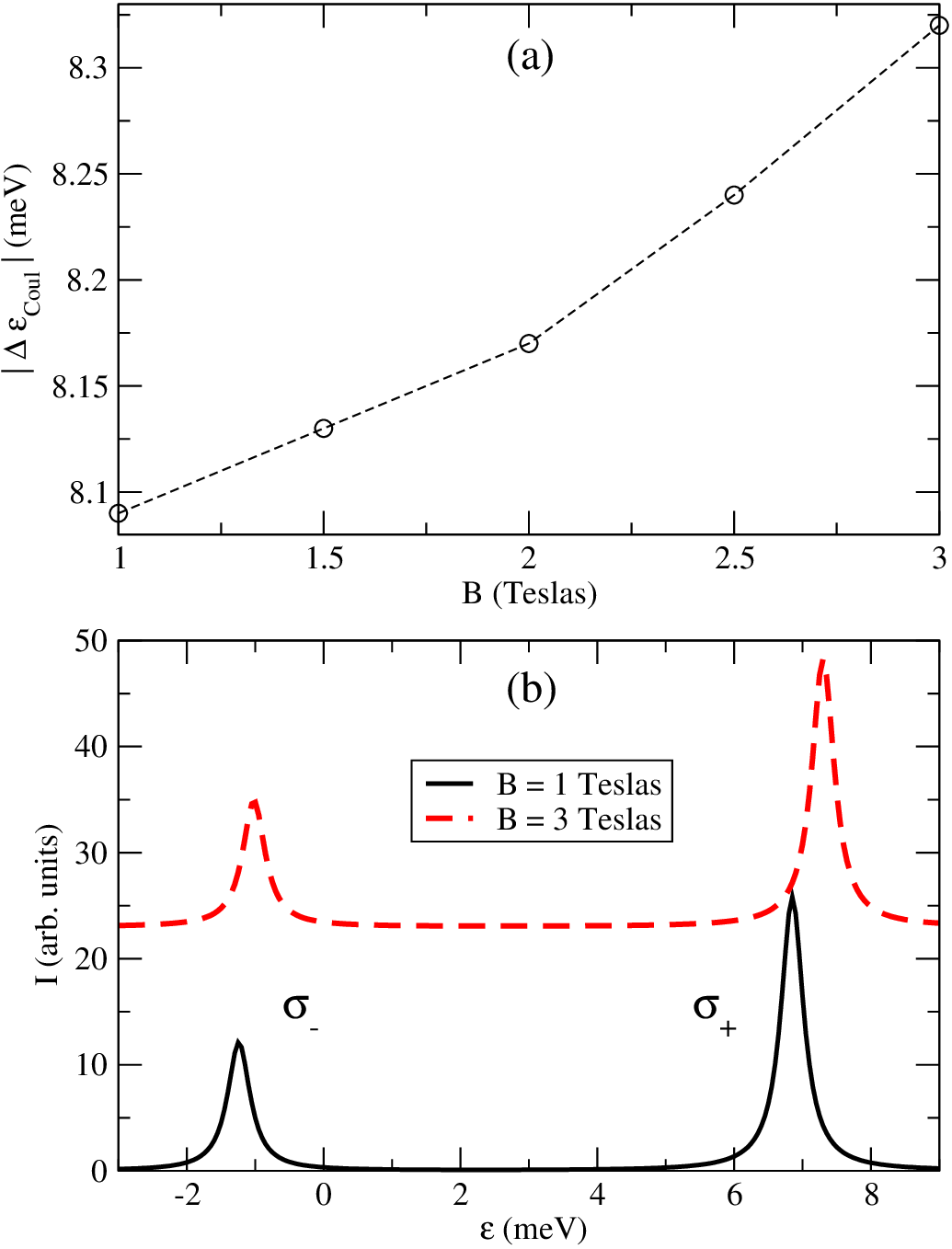}
\caption{\label{fig3} (Color online) (a) Coulomb contribution to the $\sigma_+-\sigma_-$
 splitting in the $(N_+,N_-)=(2,1)$ system, (b) Intensities of the
 luminescence lines at $B=1$ and 3 Teslas. The reference energy is $E_{gap}$.
 Only Coulomb interactions are included when computing the line positions.}
\end{center}
\end{figure}

The situation is different, however, when the number of pairs is odd. Let us
consider, for example, the $(N_+,N_-)=(2,1)$ case. We expect a nonzero mean
nuclear polarization and, consequently, a nonzero Overhauser shift. The magnitude of
this shift is expected to be around 0.1 meV. The Coulomb contribution to the
splitting is much greater, of the order of a few meV.

We show in Fig. \ref{fig3}(a) the Coulomb splitting for the luminescence lines
coming from the decay of the (2,1) system as a function of the magnetic field, $B$.
Notice the smooth dependence of $\Delta \varepsilon_{Coul}$ on $B$. We shall stress
that, when computing the position of the lines, $\varepsilon_{\pm}$, we assume
transitions from the ground state of the $(N_+,N_-)$ system to the ground
states of the $(N_+-1,N_-)$ and $(N_+,N_--1)$ systems. We verified that these
so called coherent luminescence lines account for more than 80 \% of the
total luminescence (see Appendix). We represent in Fig. \ref{fig3}(b) the relative
intensities of both lines at $B=1$ and 3 Teslas. Only Coulomb interactions are
included in the line positions. The reference energy is the band gap. Lorentzians with
a width $\Gamma=0.2$ meV are used to represent the lines. The variation of the magnetic field induces a blueshift on both line positions, which is of 0.45 meV for the $\sigma_+$ line, and 0.22 meV for the $\sigma_-$ line. For comparison, let us notice that the Zeeman energy (not included in the figure) moves the $\sigma_+$ line 0.062 meV to the left, and the $\sigma_-$ line 0.062 meV to the right, when $B$ varies from 1 to 3 Teslas.

\begin{figure}[t]
\begin{center}
\includegraphics[width=.99\linewidth,angle=0]{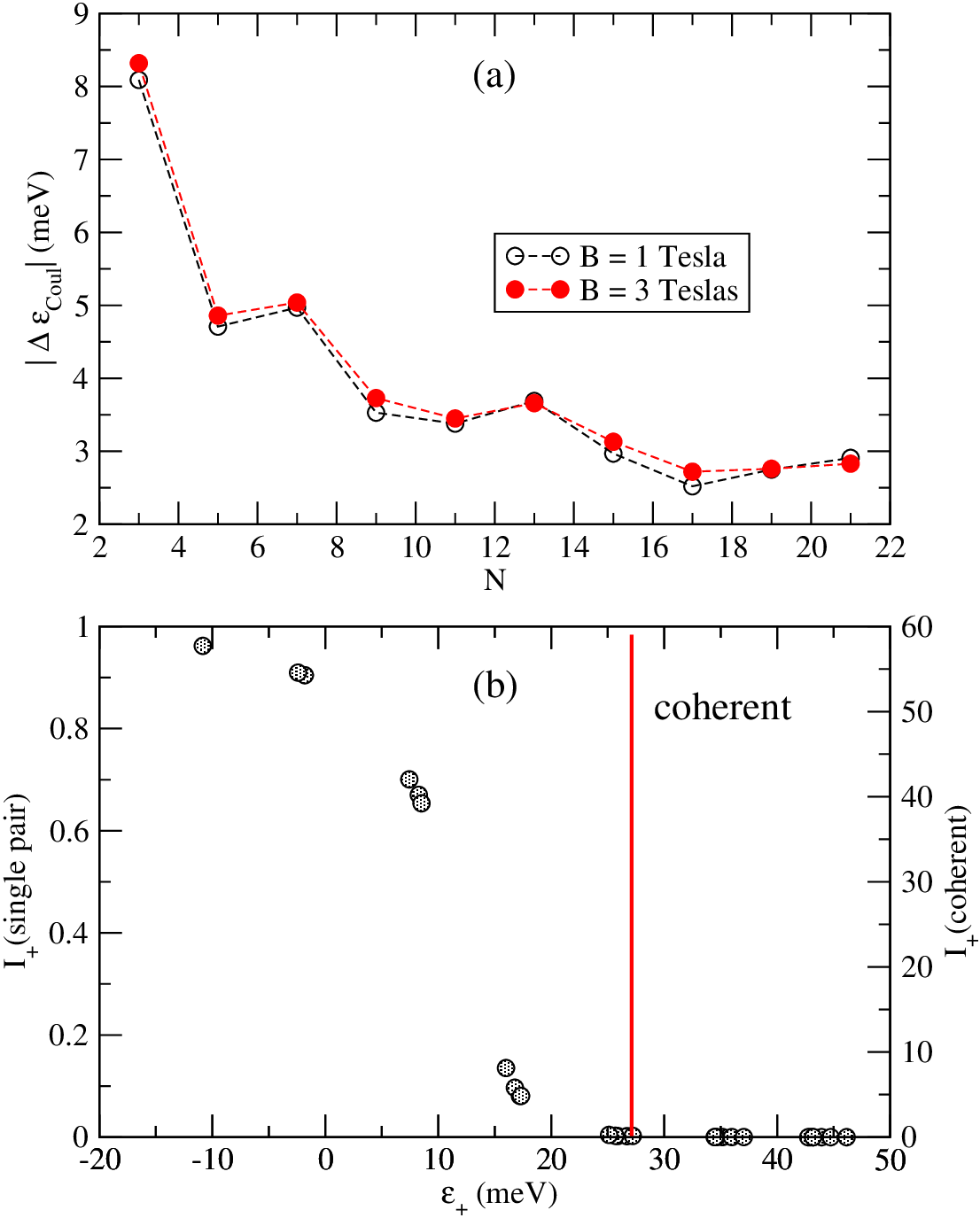}
\caption{\label{fig4} (Color online) (a) Splitting induced by Coulomb interactions
 as a function of the number of pairs in the dot, (b) Comparison between
 single-pair and coherent $\sigma_+$ luminescence in the $(N_+,N_-)=(7,6)$ system
 at $B=1$ Tesla.}
\end{center}
\end{figure}

The luminescence lines coming from the decay of the (2,1) system shall be observed
under $\sigma_-$ excitation. Due to the reasons mentioned above for the biexciton,
the excitation of the (3,0) system is hardly possible. Under $\sigma_+$ excitation,
$\Delta \varepsilon_{Coul}$ changes sign, and the lines interchange the intensities.

It is interesting to look at still higher excitations. To be definite, we consider
$\sigma_-$ pumping and follow only the luminescence lines coming from the decay of
the $(n+1,n)$ systems. The splitting induced by Coulomb interactions is drawn
in Fig. \ref{fig4}(a) as a function of the total number of pairs, $N=2 n+1$. The
splitting is close to the harmonic confinement energy, $\hbar\omega_0=10$ meV, for
the $N=3$ system, but only around 3 meV for the N=21 complex. We observe also
distinct signatures of shell filling at the magic numbers $N-1=2$, 6, 12 and 20.

The effects of Coulomb interactions are made evident in Fig.\ref{fig4}(b), where
single-pair and coherent $\sigma_+$ luminescence in the $(N_+,N_-)=(7,6)$ system
at $B=1$ Tesla are compared. The single-pair luminescence is evaluated from the
pair energies and occupations. We see the two main effects of Coulomb interactions
in this figure. First, there is a strong enhancement of luminescence intensity. This
is the analog of the well known exciton effect. Second, the coherent peak is
shifted with respect to the position of the single-pair peaks.

Finally, we shall stress that, for a multiexciton system with $N>2$ to live in a
quantum dot, the confining potential should correspond to a deep well, because these
systems are, most likely, unbound \cite{BG}. This does not seem to be the situation
reported in [\onlinecite{main},\onlinecite{main2}], where the dot is formed as a
result of fluctuations in the well width. It does not represent, however, a limitation
to the validity of our results, because quantum dots with the desired properties can
presently be grown at will.

\section{Concluding remarks}

We have shown that luminescence lines arising from the decay of one electron-hole
pair in a multiexcitonic complex may exhibit a strong asymmetry with respect to
the sense of circular polarization of the emitted light if the number of pairs
in the complex is odd. Let us stress the conditions under which this statement
is expected to be valid.

First, it was assumed that $N_+-N_-=\pm 1$. Luminescence lines from the complex
with $N_+=N_-$ were shown to exhibit roughly no Overhauser nor Coulomb shifts. From
an experimental point of view, a small disbalance between $N_+$ and $N_-$ is
expected under quasiresonant excitation and very low temperatures. High excitation
powers are needed to create the multiexcitonic complexes.

A second, very important, assumption is the coherent (collective) character of the
decay process. Lines which can be interpreted as single- or independent-pair decays
are possible too. One can guess that these single-pair decays should play an important
role in luminescence under a strong confinement regime and well above band gap
excitation\cite{multi}. In our calculations, we assumed an intermediate confinement
regime, $\hbar\omega_0=10$ meV, in which case effects due to Coulomb interactions
become very important, and the coherent luminescence accounts for more than 80 \% of
the total luminescence. The effects of Coulomb interactions were shown to be
basically the following: and enhancement of luminescence intensity and concentration
on a single line, which is shifted with respect to the position of the single-pair
decays. Band-filling in the quantum dot is seen as a blueshift of the peak position.

\begin{acknowledgments}

The authors acknowledge the Comitee for Research of the Universidad
de Antioquia for support.

\end{acknowledgments}

\appendix
\section{Lipkin-Nogami scheme with two conserved charges}
\label{appendixA}

The Hamiltonian describing the electron-hole system in the
two-dimensional quantum dot is written as:

\begin{eqnarray}
H &=& \sum_n \left(t_n^{(e)} e_n^\dagger e_n + t_n^{(h)}
 h_n^\dagger h_n\right)\nonumber\\
 &+&\frac{\beta}{2}\sum_{rsuv}
 \langle r,s|1/r|u,v\rangle ~ e_r^\dagger e_s^\dagger
 e_v e_u\nonumber\\
 &+&\frac{\beta'}{2}\sum_{rsuv}\langle r,s|1/r|u,v\rangle ~
 h_r^\dagger h_s^\dagger h_v h_u\nonumber\\
 &-&\beta''\sum_{rsuv}
 \langle r,s|1/r|u,v\rangle ~ e_r^\dagger h_s^\dagger
 h_v e_u,
 \label{eqA1}
\end{eqnarray}

\noindent
where, for simplicity, we have not included electron-hole exchange
terms. $\beta=0.8 ~ e^2/(4\pi\epsilon_0\kappa l_0)$ is the strength of
electron-electron Coulomb interactions, whereas $\beta'$ and $\beta''$
characterize the hole-hole and electron-hole interactions. The
calculations presented in the paper correspond to the symmetric
situation, in which $\beta=\beta'=\beta''$. A factor of 0.8 is included in $\beta$
to approximately account for the quasi two-dimensionality of the dot.
$l_0=\sqrt{\hbar/(m_e\Omega)}$ is the oscillator length corresponding to the modified frequency $\Omega$ (to be discussed below). The single-particle electron and hole energies in a harmonic oscillator potential and a magnetic field are given by:

\begin{equation}
t_n^{(e)}=\hbar\Omega ~(2 k+|l|+1)+\hbar\omega_c l/2,
\label{eqA2}
\end{equation}

\begin{equation}
t_n^{(h)}=\frac{m_e}{m_h}\left\{\hbar\Omega ~ (2 k+|l|+1)-
 \hbar\omega_c l/2\right\},
\label{eqA3}
\end{equation}

\noindent
where $\Omega=\sqrt{\omega^2+\omega_c^2/4}$, and $\omega_c=e B/m_e$
is the electron cyclotron frequency. It is assumed that the confinement
potential for holes is such that its characteristic length is
$l_0$ also.

$n, r, s$, etc are two-dimensional harmonic oscillator states with
characteristic length $l_0$. They are specified by the radial quantum
number, $k$, the orbital quantum number (angular momentum projection
over $B$), $l$, and the spin projection, $\sigma$. Notice that we
have not included the Zeeman and Overhauser contributions to the
single-particle energies (\ref{eqA2},\ref{eqA3}). Notice also that
the Coulomb matriz element, $\langle r,s|1/r|u,v \rangle$ is zero
unless the spin projections satisfy: $\sigma_r=\sigma_u$,
$\sigma_s=\sigma_v$.

There are four conserved charges commuting with the Hamiltonian,
(\ref{eqA1}). They are:

\begin{equation}
N_{e\uparrow}=\sum_{n\uparrow} e_n^\dagger e_n, ~~~~
 N_{e\downarrow}=\sum_{n\downarrow} e_n^\dagger e_n,
\end{equation}

\begin{equation}
N_{h\uparrow}=\sum_{n\uparrow} h_n^\dagger h_n, ~~~~
 N_{h\downarrow}=\sum_{n\downarrow} h_n^\dagger h_n,
\end{equation}

\noindent
where the notation means that only spin-up electron states enter
the sum in $N_{e\uparrow}$, etc.

In the BCS function we shall write, the number of spin-up electrons
exactly coincide with the number of spin-down holes, and the
number of spin-down electrons with the number of spin-up holes.
Let us define $N_+=N_{e\uparrow}$, and $N_-=N_{e\downarrow}$. The
mean value of the Hamiltonian, Eq. (\ref{eqA1}), should be
minimized with the constraint that the mean values
$\langle N_+\rangle$ and $\langle N_-\rangle$ should be fixed.
Mean values are obtained by averaging out with the function:

\begin{equation}
|BCS\rangle=\prod_{r\uparrow} (u_r+v_r e_r^\dagger h^\dagger_{\bar r})
 \prod_{s\downarrow} (u_s+v_s e_s^\dagger h^\dagger_{\bar s}) ~ |0\rangle,
\label{eqA6}
\end{equation}

\noindent
where $|0\rangle$ denotes the vacuum state, and the ``conjugate''
hole state $\bar n$ is defined in terms of the electron state
$n=(k,l,\sigma)$ as $\bar n=(k,-l,-\sigma)$. The coefficients
$u_r, v_r$, satisfying the normalization conditions $u_r^2+v_r^2=1$,
are to be used as variational parameters. Notice that

\begin{equation}
\langle N_+\rangle=\sum_{n\uparrow} v_n^2, ~~~~
\langle N_-\rangle=\sum_{n\downarrow} v_n^2.
\label{eqA7}
\end{equation}

The standard BCS equations (gap equations) are obtained by
introducing Lagrange multipliers $\mu_+, \mu_-$, which are
fixed from the equations (\ref{eqA7}), and minimizing the
function:

\begin{equation}
F(v)=\langle H-\mu_+ N_+ -\mu_- N_-\rangle,
\end{equation}

\noindent
with respect to the variational parameters $v_n$. The mean value
of the Hamiltonian is given by:

\begin{eqnarray}
\langle H \rangle&=&\sum_n \left( t_n^{(e)} + t_{\bar n}^{(h)}\right) v_n^2
 \nonumber\\
 &+&(\beta+\beta'-2\beta'')/2 \sum_{n,m} \langle n,m|1/r|n,m\rangle v_n^2 v_m^2
 \nonumber\\
 &-&(\beta+\beta')/2\sum_{n,m} \langle n,m|1/r|m,n\rangle v_n^2 v_m^2
 \nonumber\\
 &-&\beta''\sum_{n,m} \langle n,m|1/r|m,n\rangle v_n u_n v_m u_m).
\end{eqnarray}

\noindent
Notice the exact cancellation of direct Coulomb interactions in the
symmetric case, $\beta=\beta'=\beta''$.

A rather nontrivial fact, which helps understanding the Lipkin-Nogami
scheme, is that the BCS gap equations can also be interpreted as a
search for the optimal linear in $N_+$ and $N_-$ approximation
to $H$. Indeed, let us define $P=\lambda_0+\lambda_+ N_+ +\lambda_- N_-$, and
require the minimization of the mean square deviation:

\begin{equation}
G(\lambda)=\langle (H-P)^2\rangle.
\label{eqA9}
\end{equation}

\noindent
The equation:

\begin{equation}
\partial G/\partial \lambda_0=\langle H-P\rangle=0,
\label{eqA10a}
\end{equation}

\noindent
is used to fix the $\lambda_0$ parameter, whereas the equations $\partial G/\partial \lambda_+=\partial G/\partial \lambda_-=0$ prove to be equivalent to the BCS
gap equations. They can be written in the form:

\begin{equation}
0=\langle (H-P) N_+\rangle,
\label{eqA10}
\end{equation}

\begin{equation}
0=\langle (H-P) N_-\rangle.
\label{eqA11}
\end{equation}

The chemical potentials, $\mu_{\pm}$, can be identified with the $\lambda_{\pm}$
parameters in the present case.

In the Lipkin-Nogami method, we make a step forward and look for
a quadratic approximation to $H$:

\begin{eqnarray}
P&=&\lambda_0+\lambda_+ N_+ +\lambda_- N_- +\lambda_{++} N_+^2\nonumber\\
 &+&\lambda_{+-}N_+ N_- +\lambda_{--}N_-^2.
\end{eqnarray}

Minimization of $G$ in Eq. (\ref{eqA9}) leads, in addition to
(\ref{eqA10a},\ref{eqA10},\ref{eqA11}), to the equations:

\begin{equation}
0=\langle (H-P) N_+^2\rangle,
\label{eqA13}
\end{equation}

\begin{equation}
0=\langle (H-P) N_-^2\rangle,
\label{eqA14}
\end{equation}

\begin{equation}
0=\langle (H-P) N_+ N_-\rangle.
\label{eqA15}
\end{equation}

Eqs. (\ref{eqA10a},\ref{eqA10},\ref{eqA11},\ref{eqA13},\ref{eqA14},\ref{eqA15}) conform a linear system from which we obtain the $\lambda$ parameters in terms of averages like
$\langle H N_+^2\rangle$, $\langle N_+^2\rangle$, etc. In fact, we need only explicit
expressions for $\lambda_{++}$, $\lambda_{+-}$ and $\lambda_{--}$:

\begin{widetext}
\begin{equation}
\lambda_{+-}=\frac{\langle H N_+ N_-\rangle-\langle H N_+\rangle \langle N_-\rangle
 -\langle N_+\rangle \langle H N_-\rangle+\langle H\rangle \langle N_+\rangle
 \langle N_-\rangle}{\langle N_+^2\rangle \langle N_-^2\rangle-\langle N_+^2\rangle
 \langle N_-\rangle^2-\langle N_+\rangle^2 \langle N_-^2\rangle+\langle N_+\rangle^2
 \langle N_-\rangle^2},
\end{equation}

\begin{equation}
\lambda_{++}=\frac{(\langle H N_+^2\rangle-\langle H\rangle
 \langle N_+^2\rangle)(\langle N_+^2\rangle-\langle N_+\rangle^2)-
 (\langle H N_+\rangle-\langle H\rangle \langle N_+\rangle)
 (\langle N_+^3\rangle-\langle N_+\rangle \langle N_+^2\rangle)}{
 (\langle N_+^4\rangle-\langle N_+^2\rangle^2)(\langle N_+^2\rangle-
 \langle N_+\rangle^2)-(\langle N_+^3\rangle-\langle N_+\rangle
 \langle N_+^2\rangle)^2},
\end{equation}
\end{widetext}

\noindent
and similarly for $\lambda_{--}$. The parameters $\lambda_{+}$ and $\lambda_{-}$
are absorbed into the definition of the chemical potentials:

\begin{equation}
\mu_+=\lambda_++2 \lambda_{++} \langle N_+\rangle +
 \lambda_{+-}\langle N_-\rangle,
\end{equation}

\noindent
(and similarly for $\mu_-$). They are obtained from Eqs. (\ref{eqA7}), in which we introduce the standard parametrization:

\begin{equation}
v_{n\uparrow}^2=\frac{1}{2}\left( 1-
 \frac{\varepsilon^{HF}_{n\uparrow}-\mu_+}
 {\sqrt{(\varepsilon^{HF}_{n\uparrow}-\mu_+)^2+
 (\Delta_{n}^+)^2}}\right),
\label{eqA16}
\end{equation}

\noindent
and similarly for $v_{n\downarrow}^2$. The pair Hartree-Fock energy
is given by:

\begin{eqnarray}
\varepsilon^{HF}_{n\uparrow}&=&t_n^{(e)}+t_{\bar n}^{(h)}-\beta''\langle n,n|1/r|n,n\rangle
 -\lambda_{++} (u_n^2-v_n^2)\nonumber\\
&+&(\beta+\beta'-2\beta'')\sum_{k\uparrow\ne n\uparrow}\langle n,k|1/r|n,k\rangle v_k^2
 \nonumber\\
&+&(\beta+\beta'-2\beta'')\sum_{k\downarrow}\langle n,k|1/r|n,k\rangle v_k^2
 \nonumber\\
&-&(\beta+\beta')\sum_{k\uparrow\ne n\uparrow}\langle n,k|1/r|k,n\rangle v_k^2.
\label{eqA17}
\end{eqnarray}

\noindent
The gap parameters, $\Delta_n^+$, introduced in Eq. (\ref{eqA16}),
are obtained from Eq. (\ref{eqA10}). The latter can be put in the
form of gap equations:

\begin{equation}
\Delta_{n}^+=\beta''\sum_{s\uparrow}
 \langle n,s|1/r|s,n \rangle
 \frac{\Delta^+_{s}} {\sqrt{(\varepsilon^{HF}_{s}-\mu_+)^2+
 (\Delta_{s}^+)^2}},
 \label{eqA18}
\end{equation}

\noindent
and similarly for $\Delta_{n}^-$.

The iterative procedure designed to solve these equations is as follows.
We start from a set $\{\varepsilon_n^{(HF)}, \Delta_n\}$. Eqs. (\ref{eqA16})
are introduced into Eqs. (\ref{eqA7}) and the chemical potentials
$\mu_+, \mu_-$ are found. Then, we compute $\lambda_{++}$, $\lambda_{+-}$ and $\lambda_{--}$. Hartree-Fock energies and gap functions are recalculated from Eqs.
(\ref{eqA17},\ref{eqA18}), and the process is repeated until convergence
is reached.

For completeness, let us write the explicit expressions for the
mean values needed to compute $\lambda_{++}$, $\lambda_{+-}$ and $\lambda_{--}$:

\begin{eqnarray}
\langle N_+^2 \rangle &=& \sum_{n\uparrow,m\uparrow}\left\{
 v_n^2 v_m^2 + \delta_{nm} v_n^2 u_n^2 \right\}\nonumber\\
 &=& \langle N_+ \rangle^2 + \langle N_+ \rangle -
 \sum_{n\uparrow} v_n^4,
\end{eqnarray}

\begin{eqnarray}
\langle N_+^3 \rangle &=& \langle N_+ \rangle^3 + 3 \langle N_+ \rangle^2
 + \langle N_+ \rangle\nonumber\\
 &-& 3 \left(\langle N_+ \rangle + 1\right)
 \sum_{n\uparrow} v_n^4 + 2 \sum_{n\uparrow} v_n^6,
\end{eqnarray}

\begin{widetext}
\begin{eqnarray}
\langle N_+^4 \rangle &=& \langle N_+ \rangle^4 + 6 \langle N_+ \rangle^3
 + 7 \langle N_+ \rangle^2 + \langle N_+
 - \left( 6 \langle N_+ \rangle^2 + 18 \langle N_+ \rangle + 7\right)
 \sum_{n\uparrow} v_n^4 \nonumber\\
&+& 4 \left( 2 \langle N_+ \rangle + 3\right) \sum_{n\uparrow} v_n^6
+ 3\; (\sum_{n\uparrow} v_n^4 )^2 - 6 \sum_{n\uparrow} v_n^8,
\end{eqnarray}
\end{widetext}

\noindent
and similarly for $\langle N_-^2 \rangle$, etc. Notice that, because of
the factorizable form Eq. (\ref{eqA6}), mean values like
$\langle N_+ N_-\rangle$ factorize, $\langle N_+ N_-\rangle=
\langle N_+\rangle\langle N_-\rangle$. Concerning mean values
of $H$ with powers of $N_+$, we have:

\begin{widetext}
\begin{eqnarray}
\langle H N_+ \rangle &=& \langle H \rangle \langle N_+ \rangle +
 \sum_{n\uparrow} \left( t_n^{(e)}+t_{\bar n}^{(h)}\right) v_n^2 u_n^2
 \nonumber\\
&+&\sum_{n\uparrow,m}\left\{(\beta+\beta'-2 \beta'')\langle n,m|1/r|n,m\rangle
 -(\beta+\beta')\langle n,m|1/r|m,n\rangle\right\} v_n^2 v_m^2 (1-v_n^2)
 \nonumber\\
&-&\beta''\sum_{n\uparrow,m\uparrow}\langle n,m|1/r|m,n \rangle\;
 v_n u_n v_m u_m (1-2 v_n^2),
\end{eqnarray}

\begin{eqnarray}
\langle H N_+ N_-\rangle &=& \langle H N_+\rangle \langle N_- \rangle +
 \langle H N_-\rangle \langle N_+ \rangle -
 \langle H \rangle \langle N_+ \rangle \langle N_- \rangle\nonumber\\
&+&(\beta+\beta'-2\beta'')\sum_{n\uparrow,m\downarrow}\langle n,m|1/r|n,m\rangle\;
 v_n^2 u_n^2 v_m^2 u_m^2,
 \label{eqA28}
\end{eqnarray}

\begin{eqnarray}
\langle H N_+^2 \rangle &=& \langle H \rangle \langle N_+^2 \rangle +
 2 \langle N_+ \rangle \left(\langle H N_+ \rangle
 - \langle H \rangle\langle N_+ \rangle\right)+
 \sum_{n\uparrow} \left( t_n^{(e)}+t_{\bar n}^{(h)}\right)
 u_n^2 v_n^2 (1-2 v_n^2) \nonumber\\
&+&(\beta+\beta'-2\beta'')\sum_{n\uparrow,m}\langle n,m|1/r|n,m\rangle\;
 v_n^2 v_m^2 u_n^2 (u_n^2-v_n^2)\nonumber\\
&+&(\beta+\beta'-2\beta'')\sum_{n\uparrow,m\uparrow}\langle n,m|1/r|n,m\rangle\;
 v_n^2 v_m^2 u_n^2 u_m^2\nonumber\\
&-&(\beta+\beta')\sum_{n\uparrow,m\uparrow}\langle n,m|1/r|m,n \rangle
 (2 v_n^2 v_m^2 - 5 v_n^4 v_m^2 + 2 v_n^6 v_m^2 + v_n^4 v_m^4)\nonumber\\
&-& \beta''\sum_{n\uparrow,m\uparrow}\langle n,m|1/r|m,n \rangle\;
 v_n u_n v_m u_m (1-6 v_n^2+4 v_n^4+2 v_n^2 v_m^2).
\end{eqnarray}
\end{widetext}

Notice that, from Eq. (\ref{eqA28}) it follows that $\lambda_{+-}=0$ in the symmetric
system with $\beta=\beta'=\beta''$.
Finally, the Lipkin-Nogami energy is obtained by replacing the operators $N_+$,
$N_-$ in $P$ by its eigenvalues. It can be written in the form:

\begin{eqnarray}
E_{LN}&=&\langle H\rangle -\lambda_{++} (\langle N_+^2\rangle - \langle N_+\rangle^2)
 \nonumber\\
 &-&\lambda_{--} (\langle N_-^2\rangle - \langle N_-\rangle^2).
\end{eqnarray}

To get an idea of the intensities of the luminescence lines, we compute the
matrix elements of the interband dipole operator:

\begin{equation}
D_+= \sum_{n\uparrow} e_n h_{\bar n}.
\end{equation}

\noindent
A similar expression can be written for $D_-$. The intensity of the transition
from the ground state of the $(N_+,N_-)$ system to the ground state of the
$(N_+-1,N_-)$ system is proportional to the matrix element of $D_+$ squared:

\begin{widetext}
\begin{eqnarray}
I_+=|\langle N_+-1,N_-|D_+|N_+,N_- \rangle|^2
 = \left|\prod_{n\downarrow} (u'_n u_n+v'_n v_n)\right|^2\;
 \left|\sum_{n\uparrow} u'_n v_n \prod_{j\uparrow\ne n\uparrow}
 (u'_n u_n+v'_n v_n)\right|^2,
\end{eqnarray}
\end{widetext}

\noindent
where $(u,v)$ and $(u',v')$ are, respectively, the BCS coefficients for the
initial and final states. It can be verified that the transition from the ground state of the $(N_+,N_-)$ system to the ground state of the $(N_+-1,N_-)$ system accounts for most of the transition probability. Indeed, a total intensity can be defined in the following way:

\begin{eqnarray}
I_+^{total}&=&\sum_{\psi}|\langle \psi|D_+|N_+,N_- \rangle|^2\nonumber\\
 &=&\sum_{n\uparrow} v_n^4+(\sum_{n\uparrow} u_n v_n)^2,
\label{eqA33}
\end{eqnarray}

\noindent
where the sum runs over all possible final states, $|\psi\rangle$. For the
(2,1) complex at $B=1$ Tesla, with the parameters used in this paper, $I_+/I_+^{total}=0.84$, whereas for the (10,9) system, $I_+/I_+^{total}=0.92$.

The two terms in Eq. (\ref{eqA33}) have a simple interpretation. If pairing is
neglected, the decay of a pair in the state $n$ proceeds independently of the
rest of the states. The matrix element of $D_+$ is equal to the probability of finding
a pair in the state $n$, i.e., to $v_n^2$. Thus, $D_+$ squared equals $v_n^4$.
The first term in Eq. (\ref{eqA33}) is hence the probability of decay of individual
pairs. The second term comes from pairing. In our model, at $B=1$ Tesla, $I_+^{total}$
from the (7,6) complex, for example, has a component coming from individual decays
equal to 5.20, whereas the pairing contribution reaches the value 59.34.

\end{document}